\documentclass{article}
\begin{document}

\begin{center}
\centerline{\large \bf Entanglement and inequality
of forward and reversed processes}
\centerline{\large \bf as a physical base of nonlinear phenomena in optics.}
\end{center}

\vspace{3 pt}
\centerline{\sl V.A.Kuz'menko\footnote{Electronic 
address: kuzmenko@triniti.ru}}

\vspace{5 pt}
\centerline{\small \it Troitsk Institute for Innovation and Fusion 
Research,}
\centerline{\small \it Troitsk, Moscow region, 142190, Russian 
Federation.}

\vspace{5 pt}
\begin{abstract}

	Connection between the concept of entanglement and origin of nonlinear 
phenomena in optics is discussed.

\vspace{5 pt}
\end{abstract}

\vspace{12 pt}

 The concept of entanglement plays an important role in quantum mechanics 
and usually it is connected with so-called nonlocality of quantum mechanical 
interactions \emph{in space} [1, 2]. Many years it is extensively studied 
due to fundamental interest and also for possible practical applications 
in the fields of communication and quantum computation. Substantially 
less attention is devoted to the so-called \emph{entanglement in time} [3-7].
      
	The concept of \emph{entanglement in time} can also have other sense 
as a \emph{memory} of quantum system about the \emph{initial state}. 
Does such memory possible? \emph {The experiments clearly show that such memory 
of quantum system about the initial state really exists and manifests 
itself through inequality of forward and reversed processes in optics} 
[8-10]. Efficiency of the reversed process may be much greater, 
than the efficiency of forward one. 

	In the recent experimental works [8, 9] the forward (splitting) 
and reversed (mixing) processes with a photons were studied (Fig.1b). 
On the first stage the narrowband (0.04 nm) radiation of nanosecond 
laser was transformed through down-conversion in the nonlinear crystal 
into two intense broadband beams (each spectral width $ \sim 100 nm $). 
On the second stage this two broadband beams were mixed in the sum 
frequency generator [9] or in the process of two photon excitation 
of rubidium atoms [8]. The mixing of entangled photons leads to 
regeneration of initial narrowband radiation and is the example of 
reversed process into the \emph{initial state}. In contrast, the mixing of 
non-entangled photons should give broadband radiation and is the example 
of again only forward process. Both experiments show the same result: 
the efficiency of reversed process is much greater, than the efficiency of 
forward process. In these experiments the entangled photons demonstrate 
the \emph{memory} about the \emph{initial state}. 

	In the next case a polyatomic molecules demonstrate such memory [10]. 
Excitation and deexcitation of $ SF_6$ molecules by $ CO_2$-laser radiation 
was studied in this work (Fig.1a). The forward process (photon absorption) 
in this case has extremely high spectral width $\sim 150 $ GHz (the so-called 
wide component of line [11]). In contrast, the reversed process (stimulated 
emission) has very small spectral width. The difference in spectral width 
of forward and reversed processes exceeds five orders of magnitude. 
Accordingly, the cross-section of the reversed process turns out to be 
in several orders of magnitude greater, than the cross-section of forward 
process [12]. 

	The discussed experiments [8-10] are the direct experimental proofs 
of inequality of forward and reversed processes. Furthermore, we have 
enormous quantity of indirect proofs. The nonlinear optics, as a whole, looks 
like as a large accumulation of indirect evidences of such inequality. 
Nonlinear phenomena in optics usually are described by mathematical models 
which are based on the Bloch equations. Such mathematical models usually 
give good description of nonlinear effects but, in fact, its do not have 
any clear physical sense. The concepts of coherent states, wavepackets and 
their interference are used for explanation of physical nature of nonlinear 
phenomena. At the same time the question about the existence of coherent 
states is debated in literature [13]. Some theorists argue that the 
inability to measure the absolute phase of an electromagnetic field 
prohibits the representation of a lasers output as a quantum optical 
coherent state [14, 15]. In fact, it means, that using the concept of 
coherent states for explanation of nonlinear phenomena does not have 
physical sense. 

	Alternative explanation of origin of nonlinear phenomena in optics 
is based on the concept of inequality of forward and reversed processes [16].
It gives really simple and clear interpretation of origin of nonlinear 
phenomena. A terminal stage of nonlinear processes (there are usually a 
multiple step photon mixing processes) is returning the system into the 
initial state. This process proceeds with extremely high efficiency. 
Because of a photon has spin, it is impossible to return exactly into the 
initial state with odd number of photons. That is why in the gas and liquid 
phase the observed nonlinear effects are based only on schemes with even 
number of photons (four, six and so on photon mixing). In a solid state 
with rigid lattice the rotational motion of atoms and molecules may be 
completely suppressed and the role of a photon's spin vanishes. In this 
case the nonlinear effects with odd number of photons become possible (three 
photon mixing). 

	The concept of \emph{initial state} of quantum system should include 
also the orientation of molecule in space and even the phase of vibrational 
motion of atoms. Thanks to this fact the experimental study of rotational 
and vibrational motions of molecules in the real time become possible 
[17, 18]. 

	So, we need to originate a new mathematical model (alternative to 
the Bloch equations) for description the dynamics of optical transitions. 
Such model should be based on the concept of inequality of forward and 
reversed processes. 

	In conclusion, we point out that the study of 
\emph{entanglement in time} does not less important task, than the study 
of \emph{entanglement in space}.

\end{document}